\documentstyle[12pt]{article}
\newcommand{\Rset}[0]{\mbox{\rm I\kern-.200em R}}

\title{THE THEORY OF IDEAL YANG-MILLS
FLUIDS IN SYMMETRIC HYPERBOLIC FORM
\thanks{This work
has been supported by NSF grant PHY 84-04035.}}
\author{Maurice H.P.M. van Putten\\
	Institute for Theoretical Physics\\
        UCSB, Santa Barbara, CA 93106, USA}
\date{}

\begin{document}

\maketitle

\begin{abstract}
The theory describing the evolution of
perfect fluids coupled to a Yang-Mills field in the
presence of infinite electrical conductivity
(ideal Yang-Mills fluids) is proved to be hyperbolic.
This result is obtained by embedding the
theory in a symmetric hyperbolic system of equations.

A divergence formulation of the theory of
ideal Yang-Mills fluids forms the basis of our analysis.
This divergence form constitutes a covariant
and constraint-free formulation.
The infinitesimally small amplitude
wave structure is derived, and found to be determined
by a sixth-order polynomial.
Alfven waves ($\delta P=\delta r=0$) do not, in general,
exist, and occur in very special cases only.
The theory is embedded in a symmetric hyperbolic
system using a generalization of Friedrichs
symmetrization procedure. This establishes
at once both algebraic hyperbolicity (all wave speeds
real and less than unit) and
dependence on initial data to within
finite order.
\end{abstract}

\bibliographystyle{plain}

\baselineskip=22pt

\section{Introduction}
\label{sc1}

A theory of Yang-Mills fluids
has recently been discussed in the limit
of infinite conductivity, {\em ideal Yang-Mills fluids} (IYMF),
by Choquet-Bruhat \cite{cb:91,cb:92}.
This describes relativistic fluids with a colored magnetic fields.
In this paper, we concentrate on the mathematical structure of
such Yang-Mills fluids.
The main purpose is to establish
well-posedness of initial value problems in this theory. Specifically,
it will be established

\newtheorem{TH1}{Theorem}
\begin{TH1}
Consider relativistic hydrodynamics
in which initial value problems
are well posed by suitable equation of state.
For a fluid with infinite conductivity,
well-posedness is preserved in the
presence of a Yang-Mills field.
\end{TH1}

It is unknown whether a field description for
colored magnetic fields is physically appropriate to describe the
dynamics of high density matter at finite temperature,
in view of our uncertainty about screening of colored magnetic fields.
The well-posedness result Theorem 1 above shows that
weak screening of colored magnetic fields, though uncertain, is permissable.

We also derive expressions for the
infinitesimally small amplitude wave structure.
While finite conductivity decouples
the hydrodynamical and ``electromagnetic waves,"
this is no longer true in the singular limit of
infinite conductivity.

The analysis is based on two reformulations of the theory of IYMF.
We begin by obtaining the theory of ideal
Yang-Mills fluids
in {\em divergence form}
following \cite{mvp:91,mvp:93b}. This formulation is fully
covariant and constraint-free. It contains the Yang-Mills
solutions by dynamically conserving the original flux-freezing
constraints. This formulation
Cauchy-Kowalewski regularizes the original formulation of IYMF
($cf.$ \cite{cb:a}) and, therefore, defines
the characteristic determinant.
Continuing, we establish well posedness through symmetrization.
Friedrichs \& Lax \cite{fr:71} and Friedrichs \cite{fr:74} introduced
a symmetrization procedure for systems of conservation laws
in the presence of a ``main dependency relation" and
a certain convexity property.
In this paper, a generalization of their symmetrization procedure
is presented which is appropriate for the divergence formulation
of ideal Yang-Mills fluids. This symmetrization
procedure allows for the
theory of ideal Yang-Mills fluids to be embedded
in symmetric hyperbolic form.
Given this, the standard regularity results for
quasi-linear symmetric hyperbolic systems apply
(see, in particular, Fisher \& Marsden \cite{fm:72} and referenced
therein),
which establishes dependence on initial data to within finite order.

The organisation of this paper is as follows.
In Section 2, the theory of ideal Yang-Mills fluids with
infinite conductivity is stated. This theory is formulated in
covariant, constraint-free divergence form in Section 3.
The characteristic determinant
is derived in Section 4.
A symmetrization procedure in the presence of conserved
constraints is presented in Section 5.
Using this symmetrization procedure,
ideal Yang-Mills in divergence form is shown to
allow for an embedding in a symmetric hyperbolic system
(Section 6).

\section{Ideal Yang-Mills fluids}
\label{sc2}
The theory of Yang-Mills fluids describes the
evolution of a compressible fluid
possessing a magnetic field, $h_a$, with values in a non-abelian
Lie algebra. In the presence
of infinite conductivity,
the magnetic field is governed by the Yang-Mills'
generalization of Faraday's induction law.
This description may be given in the presence of a
curved space-time background with prescribed
hyperbolic metric $g_{ab}$ (thereby considering
Yang-Mills $test$ fluids). We will do so
with metric signature $(-,+,+,+)$.

The fluid to be considered is assumed
to be inviscid and compressible with given equation of state.
For example, the fluid may be modelled with a
polytropic equation of state, relating the
isotropic hydrostatic
pressure, $P$, to the rest mass density, $r$, as
\begin{eqnarray}
P=Kr^{\gamma}.
\label{peqs}
\end{eqnarray}
Here, $K$ is the adiabatic constant
and $\gamma$ is the polytropic index.
$K$ is conserved along the fluid's world lines,
except across shocks.
For a discussion of thermodynamics of relativistic
fluids in some generality, see Lichnerowicz \cite{li:b}.
Let the velocity four-vector field $u^a$ represent the tangent to
the world lines of the fluid elements.
Then $U=(u^a,r,P)$ decribes the state of the fluid
elements, and is governed by conservation of energy-momentum
and baryon number:
\begin{eqnarray}
\left\{
\begin{array}{l}
\nabla_aT^{ab}_f(U)=0,\\
\nabla_a(ru^a)=0,\\
c_0(U):=u^2+1=0.
\end{array}
\right.
\label{HD}
\end{eqnarray}
Here, $T_f^{ab}=rfu^au^b+Pg^{ab}$ is the stress-energy
tensor for a perfect fluid with
the specific enthalpy,
$f$. For example, $f:=1+\frac{\gamma}{\gamma-1}\frac{P}{r}$
in case of a polytropic equation of state (\ref{peqs}). Notice that
(\ref{HD}) constitutes six equations in the six unknowns $U$.

The theory of Yang-Mills fluids in the presence
of infinite conductivity describes the dynamic interation
between a perfect fluid and a (charged) magnetic field given by a
one-form, $h_a$, with its
values in a non-abelian Lie algebra ${\cal G}$ of dimension $N$.
${\cal G}$ possesses
totally antisymmetric
structure constants $c_{\lambda\mu\nu}$ (see, $e.g.$,
\cite{kh:82}).
Elements from
${\cal G}$ will be indexed in Greek.
We thus consider a problem in $6+4N$ variables
$U=(u^a,h_a,r,P)$. To be precise,
the Yang-Mills ${\cal G}$-valued 2-form
\begin{eqnarray}
F_{ab}:=\nabla_aA_b-\nabla_bA_a+[A_a,A_b]
\label{defF}
\end{eqnarray}
in the presence of infinite conductivity,
\begin{eqnarray}
u^aF_{ab}\equiv0,
\end{eqnarray}
may be expressed
in terms of a ${\cal G}$-valued 1-form $h_a$ as
\begin{eqnarray}
F_{ab}=\epsilon_{abcd}h^cu^d.
\label{YMF}
\end{eqnarray}
Here, $\epsilon_{abcd}$ is the
natural volume element on the space-time manifold.
The condition $u^ch_c=0$ further imposes uniqueness on this
representation (\ref{YMF}) of the Yang-Mills field.
The evolution of $h_a$ is governed by the so-called
{\em Yang-Mills equations}
(see \cite{cb:91,cb:92} for a
derivation)
\begin{eqnarray}
K:
\left\{
\begin{array}{l}
\hat{\nabla}_a\{h^{[a}u^{b]}\}=0,\\
c(U):=h^cu_c=0.
\end{array}
\right.
\label{KYM}
\end{eqnarray}
Here, the gauge invariant derivative
$\hat{\nabla}_a:=\nabla+[A,\cdot]$ is introduced
so as to ensure gauge invariance in (\ref{KYM})
($cf.$ \cite{cb:91,cb:92}).
The conditions $c_\lambda(U):=h^c_\lambda u_c=0$ constitute
$N$ constraints, one on each Lie-element $h_\lambda^a$,
and may be considered the
{\em flux-freezing} constraints in Yang-Mills fluids following
the nomenclature in ideal magneto-hydrodynamics.
They impose the condition that the magnetic field
be purely spatial in the fluid's rest frame.

The magnetic field-fluid coupling is established
by nonzero magnetic field energy and stresses.
In the limit of infinite conductivity, this amounts
to adding the stress-energy tensor $T_m^{ab}$,
\begin{eqnarray}
T^{ab}_m(U)=h^2u^au^b+\frac{1}{2}h^2g^{ab}-h^ah^b,
\end{eqnarray}
to the fluid
stress-energy tensor, $T_f^{ab}(U)$, mentioned above.
Here, summation over the Lie-index is implicit.
In anticipation of what follows, it becomes appropriate
to define the {\em Yang-Mills magnetic stress tensor}
$H^a_b:=h^ah_b\equiv
\sum_\lambda h^b_\lambda h_{\lambda a}$. Notice that
$T^{ab}_m(U)$ is entirely determined by $H^b_a$,
as $h^2=H^c_c$.
The evolution of Yang-Mills fluids in the presence of
infinite conductivity as a test fluid
is now
described by \cite{cb:91,cb:92}
\begin{eqnarray}
\left\{
\begin{array}{c}
\nabla_a T^{ab} = 0,\\
\hat{\nabla}_a \{h^{[a}u^{b]}\} = 0,\\
\nabla_a (ru^a) = 0,\\
c_\lambda(U) = 0,\\
c_0(U) = 0,
\end{array}
\right.
\label{PDAE}
\end{eqnarray}
where $T^{ab}(U)=T^{ab}_f(U)+T^{ab}_m(U)$ is the total stress-energy tensor.
Notice that (\ref{PDAE}) constitutes essentially $6+4N$ equations.
Of course, (\ref{PDAE}) must be closed by (\ref{defF}) and (\ref{YMF}).

It is imperative to observe
that (\ref{PDAE}) constitutes a
partial differential-algebraic system of equations.
Without the constraints $c_\lambda(U)=c_0(U)=0$ the remaining
partial differential equations in (\ref{PDAE}) do not impose uniqueness
on the evolution of the unknowns $U$.
In other words, (\ref{PDAE}) is not regular in the sense of Cauchy-Kowalewski
($cf.$ \cite{cb:a}). On the other hand, systems of purely partial-differential
equations are the most commonly studied, both analytically and
numerically.
It now becomes of
interest to first obtain the theory of ideal Yang-Mills fluids
in covariant form
as a system of precisely this type with no constraints.

In what follows we will refer to the theory described by
(\ref{PDAE}) as the theory of $ideal$ Yang-Mills fluids, analoguous
to the nomenclature $ideal$ magneto-hydrodynamics in the case
of a single magnetic field ($N=1$) with infinite `electrical'
conductivity.

\section{Ideal Yang-Mills in divergence form}
\label{sc3}
Consider the initial value problem for ideal Yang-Mills fluids.
Maxwell's
equations require initial data, $U^{(0)}$, to satisfy
certain compatibility conditions.
Let $\Sigma$ denote an initial, space-like
submanifold with (time-like) normal one-form, $\nu_a$, and
write
\begin{eqnarray}
\hat{\nabla}_a=-\nu_a (\nu_c\hat{\nabla}^c)+(\hat{\nabla}_\Sigma)_a
\end{eqnarray}
for $\hat{\nabla}_a$ on $\Sigma$. The initial data $U^{(0)}$ are then
subjected to the conditions ($cf.$ \cite{mvp:91} for $N=1$)
\begin{eqnarray}
\left\{
\begin{array}{l}
\nu_b(\hat{\nabla}_{\Sigma})_a\{h^{[a}u^{b]}\}=0,\\
c_\lambda(U)=0.
\end{array}
\right.
\label{COMP}
\end{eqnarray}
This constitutes a covariant formulation of the
condition that the initial magnetic field be divergence free
in the space-like, $\nu_a-$orthogonal submanifold $\Sigma$, and
that the magnetic field is purely space-like for observers
whose world-lines have tangents $u^a$.

Application of the technique described in \cite{mvp:91}
to each $K_\lambda$, $\lambda=1,\cdots,N$,
yields \cite{mvp:ITP}
\newtheorem{mhd}{Theorem}[section]
\begin{mhd}
The equations of ideal Yang-Mills
fluids including the flux-freezing
constraints can be stated as
\begin{eqnarray}
\hat{\nabla}_aF^{aA} \equiv
\left\{
\begin{array}{c}
\nabla_a T^{ab} = 0,\\
\hat{\nabla}_a \{h^{[a}u^{b]}+g^{ab}c(U)\} = 0,\\
\nabla_a (ru^a) = 0,\\
\nabla_a \{(u^cu_c+1)\xi^a\} = 0.
\label{PDE}
\end{array}
\right.
\end{eqnarray}
This system must be closed by
\begin{eqnarray}
\nabla_aA_b-\nabla_bA_a
=\epsilon_{abcd}h^cu^d
-[A_a,A_b].
\label{closure}
\end{eqnarray}
The vector $\xi^a$ is any prescribed time-like vector field. This system
is equivalent to (\ref{PDAE}) in regions where the flow is continuously
differentiable. The standard jump conditions across surfaces of
discontinuity for this system
are those of conservation of energy-momentum, baryon number
and Yang-Mills equations.
\end{mhd}

$Proof:$ Clearly, we need only show that a solution to an
initial value problem in the
new formulation with Cauchy-data
satisfying compatibity conditions (\ref{COMP})
on an initial, space-like hypersurface $\Sigma$,
yields a solution to the original system of
PDAE's (\ref{PDAE}). We will do so by showing that $c_\lambda$
satisfies the a homogeneous wave equation
with vanishing Cauchy-data:
\begin{eqnarray}
\left\{
\begin{array}{ll}
\Box c=0\mbox{      }&\mbox{in  }D^+(\Sigma),\\
c=0&\mbox{on   }\Sigma,\\
\nu^a\nabla_ac=0&\mbox{on   }\Sigma.
\end{array}
\right.
\end{eqnarray}
Here, we have defined the
{\em Yang-Mills wave operator }
$\Box=\nabla^c\nabla_c+[(\nabla^cA_c),\cdot]
+[A^c\nabla_a,\cdot]+[A^c,[A_c,\cdot]]$.
$D^+(\Sigma)$ denotes the future domain of dependence
of $\Sigma$, and $\nu_a$ denotes the normal one-form
to $\Sigma$. In what follows,
$\epsilon_{abcd}=\sqrt{-g}[abcd]$
will denote the volume element on our four-dimensional
manifold with $\sqrt{-g}=\mbox{det}[g_{ab}]$. Here,
$[abcd]$ is the totally antisymmetric symbol such that
$[0123]=1$.\mbox{}\\
\mbox{}\\
Ricci's identity implies the identity
\cite{cb:92}
\begin{eqnarray}
\hat{\nabla}_a\hat{\nabla}_b(2h^{[a}u^{b]})
+\frac{1}{4}[F_{ab},F_{cd}-\epsilon_{cdef}h^eu^f]\epsilon^{abcd}
\equiv0.
\end{eqnarray}
By (\ref{closure}),
$F_{ab}=\epsilon_{abcd}h^cu^d$ is enforced,
so that $\hat{\nabla}_a\hat{\nabla}_b(2h^{[a}u^{b]})\equiv0$.
Consequently,
we are left with the Yang-Mills wave equation for $c$:
\begin{eqnarray}
0=\hat{\nabla}_a\hat{\nabla}_b(\omega^{ab})=
\hat{\nabla}_a\hat{\nabla}_b(h^{[a}u^{b]})
+(\hat{\nabla}^c\hat{\nabla}_c)c=
\Box c.
\end{eqnarray}
On $\Sigma$, we may write
\begin{eqnarray}
\hat{\nabla}_a=-\nu_a(\nu^c\hat{\nabla}_c)+(\hat{\nabla}_\Sigma)_a.
\end{eqnarray}
Using this, Cauchy-data satisfying (\ref{COMP}) yield
\begin{eqnarray}
\begin{array}{ll}
0&=\nu^b\{\hat{\nabla}^a(h_{[a}u_{b]}+g_{ab}c)\}\\
&=-\nu^b\nu^a(\nu^c\hat{\nabla}_c)\{h_{[a}u_{b]}\}
+\nabla^b(\hat{\nabla}_\Sigma)^a\{h_{[a}u_{b]}\}
+\nu^b\hat{\nabla}_bc\\
&=\nu^b\nabla_bc,
\end{array}
\end{eqnarray}
because $h_{[a}u_{b]}$ is
antisymmetric.

This forces $c\equiv0$ in $D^+(\Sigma)$,
and the proof is complete. $\Box$

We remark
that for isentropic fluids, where the entropy is constant everywhere,
the fluid variables $r$ and $P$ depend on essentially one parameter,
in view of $\mbox{d}P|_{S}=r\mbox{d}f$.
In this event, conservation
of energy-momentum, $\nabla_aT^{ab}(U)=0$, together with conservation of
baryon number, $\nabla_a(ru^a)=0$, may be seen to yield conservation of the
constraint
$c_0(U)=0.$
(\ref{PDE}) then reduces to
\begin{eqnarray}
\left\{
\begin{array}{c}
\nabla_a T^{ab} = 0,\\
\hat{\nabla}_a \{h^{[a}u^{b]}+g^{ab}c(U)\} = 0,\\
\nabla_a (ru^a) = 0.
\end{array}
\right.
\label{PDAE0}
\end{eqnarray}

(\ref{PDE}) strictly contains
ideal Yang-Mills solutions to PDAE (\ref{PDAE}).
Ideal Yang-Mills solutions
are characterized by vanishing of the flux-freezing constraints
\begin{eqnarray}
c_\lambda(U)=0.
\end{eqnarray}
Initial data satisfying the compatibility conditions
(\ref{COMP}) ensure that the $c_\lambda(U)$ are preserved as zero
(as in the case of MHD \cite{mvp:91}).
The formulation of the constraint $c_0(U)$ in terms of
the last equation in (\ref{PDE}) is taken from the
divergence formulation of ideal MHD \cite{mvp:93b}.

\section{Ideal Yang-Mills' characteristic determinant}
\label{sc4}
Infinitesimally small amplitude waves may be defined through
the normal cone, ${\cal N}$, of the
one-forms $\nu_a$ normal to their time-like surfaces
of propagation. We will consider waves whose
velocity of propagation is strictly less than unity,
$i.e.$, waves with $\nu^c\nu_c\ne0$.
The condition on $\nu_a\epsilon{\cal N}$ is given
by the vanishing of the characteristic determinant
(pointwise on the space-time manifold)
\begin{eqnarray}
D(U;\nu_a):=\mbox{det}
\frac{\partial F^{aA}\nu_a}{\partial U^B} =0.
\label{YMCD}
\end{eqnarray}
Here, $B$ indexes the unknowns $U=(u^a,h_a,P,r)$, as defined before.
Clearly, $D(U;\nu_a)$ constitutes a
$6+4N$ degree polynomial in $\nu_a$. The wave structure may be
classified by finding the factorization of $D(U;\nu_a)$ over
miminal polynomials in $\nu_a$.

Using Theorem 3.1, we have
\begin{eqnarray}
D(U;\nu_a)=\mbox{det}\left(
\begin{array}{cccccc}
K 	& R_1    & \cdots &           	& R_N & | * \\
\hline\\
S^1 	& A      &        &     	&     & | 0 \\
S^2 	& 	 & A      &     	&     & | 0 \\
\cdots 	&        &        & \cdots    	&     &     \\
S^N 	&        &        & 		& A   & | 0 \\
\hline\\
\mbox{*}& 0      & \cdots &             & 0   & | *
\end{array}
\right).
\label{YMCD1}
\end{eqnarray}
Here, we have introduced the 4 by 4 matrices (when written in component
form)
\begin{eqnarray}
\begin{array}{l}
K^b_a:=\frac{\partial T^{cb}\nu_c}{\partial u^a}
=(rf+h^2)(u^c\nu_c)\delta^b_a+(rf+h^2)u^b\nu_a,\\
R^b_{\lambda a}:=\frac{\partial T^{cb}\nu_c}{\partial h_\lambda^a}
=2(u^c\nu_c)u^bh_{\lambda a}+\nu^bh_{\lambda a}-h^b_\lambda\nu_a
-(h^c_\lambda\nu_c)\delta^b_a,\\
S^{\lambda b}_a:=\frac{\partial\omega^{cb}_\lambda\nu_a}{\partial u^a}
=-h^b_\lambda\nu_a+(h^c_\lambda\nu_c)\delta^b_a+\nu^bh_{\lambda a},\\
A^b_a:=\frac{\partial\omega^{cb}_\lambda\nu_c}{\partial h^a_\lambda}
=-(u^c\nu_c)\delta^b_a+u^b\nu_a+\nu^bu_a,
\end{array}
\label{KRSA}
\end{eqnarray}
where
$\omega^{ab}_\lambda:=h^au^b_\lambda-h^bu^a_\lambda+g^{ab}c_\lambda(U)$.
The main step towards factorization is contained in

\newtheorem{ML}{Proposition}[section]
\begin{ML}
The Yang-Mills characteristic determinant $D(U;\nu_a)$
contains the factor $|A|^{N-1}$.
\end{ML}

The proof of the Proposition is based on a suitable rewriting of the
characteristic matrix $[F^B_A](U;\nu_a)$ in
(\ref{YMCD}) and (\ref{YMCD1}), and the observation
\newtheorem{l1}{Lemma}[section]
\begin{l1}
Consider a rank-one update, $M_\lambda$,
to an $N\times N$ matrix, $M=[m^b_a]$, through
an $N$-dimensional column
vector, $u^b$, and row vector, $v_a$, of the form
$M+\lambda [u^bv_a]$.
The determinant of
$M_\lambda=M+\lambda [u^bv_a]$ is linear in $\lambda$.
\end{l1}

$Proof:$ If either $u^b$ or $v_a$ is zero, there is nothing to prove.
Let, therefore, $v_k\ne0$. We have
\begin{eqnarray}
\begin{array}{lll}
\mbox{det } M_\lambda&=&
\mbox{det } [[m^b_1+\lambda u^bv_1],\cdots,[m^b_N+\lambda u^bv_N]]\\
&=&
\mbox{det } [[m^b_1-\frac{v_1}{v_k}m^b_k],\cdots,
[m^b_{k-1}+\lambda u^bv_{k-1}],
[m^b_k+\lambda u^bv_k],\\
& &
[m^b_{k+1}+\lambda u^bv_{k+1}],\cdots,
[m^b_N-\frac{v_N}{v_k}m^b_N]]\\
&\equiv&
\mbox{det } [[\bar{m}^b_1],\cdots,[\bar{m}^b_{k-1}],
[m^b_k+\lambda u^bv_k],[\bar{m}^b_{k+1}],
\cdots,[\bar{m}^b_N]].
\end{array}
\end{eqnarray}
Because of the definition
$\mbox{det }[H^b_a]=
\sum_\sigma (-1)^{\epsilon_{\sigma_1\cdots\sigma_N}}
H^1_{\sigma_1}H^2_{\sigma_2}\cdots H^N_{\sigma_N}$ for the
determinant of an $N\times N$ matrix $[H^b_a]$,
where $\epsilon$ is the $N-$dimensional totally antisymmetric
symbol and the sum is over all $N$-permutations $\sigma$,
it follows that each term in our determinant
contains $\lambda^k$ with $k=0,1$. $\Box$
\mbox{}\\
\mbox{}\\
{\em Proof of Proposition 4.1:}
The result will follow in several steps. The first two steps
concern rewriting of the characteristic matrix of Yang-Mills'
in divergence form through column block operations
by (unitary matrix) multiplications
from the $right$ and row block operations by (unitary matrix)
multiplications from the
$left$. The final result follows from further inspection of a
resulting 6 by 6 reduced matrix.

\underline{Step (a)}: Operate on the leading column
block through multiplication by matrices with unit determinant
from the $right$:
\begin{eqnarray}
D(U;\nu_a)=
[\frac{\partial F^{aA}\nu_a}{\partial U^B}]
\left(
\begin{array}{cccccc}
\mbox{id}& 0	 & 0	  & \cdots     	& 0 & | 0 \\
\hline\\
-A^{-1}S^1&\mbox{id}&        &     	&     & | 0 \\
0 	& 	 &\mbox{id} &     	&     & | 0 \\
\cdots 	&        &        & \cdots    	&     &     \\
0 	&        &        & 		&\mbox{id}& | 0 \\
\hline\\
0	& 0      & \cdots &             & 0   & |\mbox{id}
\end{array}
\right)
\end{eqnarray}

\begin{eqnarray}
=\mbox{det}\left(
\begin{array}{cccccc}
K-R_1A^{-1}S^1& R_1   & \cdots &           	& R_N & | * \\
\hline\\
0 	& A      &        &     	&     & | 0 \\
S^2 	& 	 & A      &     	&     & | 0 \\
\cdots 	&        &        & \cdots    	&     &     \\
S^N 	&        &        & 		& A   & | 0 \\
\hline\\
\mbox{*}& 0      & \cdots &             & 0   & | *
\end{array}
\right).
\label{CD1}
\end{eqnarray}

Proceeding in this manner to cancel each $S^\lambda$ in the leading
column block through subsequent multiplication of the matrix in
(\ref{CD1}) from the right by matrices with unit determinant of the form
\begin{eqnarray}
\left(
\begin{array}{cccccccc}
\mbox{id}& 0	 &\cdots 	  & & & &    	&  | 0 \\
\hline\\
0	 &\mbox{id}&      & & &     		&   &| 0 \\
\cdots 	         & &         & \cdots  &         & & & \\
0	         & &\mbox{id}&         &         & & &| 0 \\
-A^{-1}S^\lambda & &         &\mbox{id}&         & & &| 0 \\
0	         & &         &         &\mbox{id}& & &| 0 \\
\cdots 	&        &        & \cdots    	&         & & & \\
0 	&        &        & 		&        &\mbox{id}&&|0\\
\hline\\
0	& \cdots & & & &             & & |\mbox{id}
\end{array}
\right),
\end{eqnarray}
we obtain
\begin{eqnarray}
D=\mbox{det}\left(
\begin{array}{cccccc}
K-RA^{-1}S& R_1    & \cdots &           	& R_N & | * \\
\hline\\
0 	& A      &        &     	&     & | 0 \\
0 	& 	 & A      &     	&     & | 0 \\
\cdots 	&        &        & \cdots    	&     &     \\
0 	&        &        & 		& A   & | 0 \\
\hline\\
\mbox{*}& 0      & \cdots &             & 0   & | *
\end{array}
\right)
\equiv \mbox{det}
[\frac{\partial\bar{F}^{aA}\nu_a}{\partial U^B}].
\end{eqnarray}
Here, we have used
\begin{eqnarray}
RA^{-1}S\equiv\sum_\lambda R_\lambda A^{-1}S^\lambda.
\end{eqnarray}

\underline{Step (b):}
By a procedure similar to that used in Step (a),
but now row-wise, we may proceed to
cancel each $R_\lambda$ in the leading row block through multiplication by
matrices with unit determinant from the $left$. Beginning with cancellaetion of
$R_1$,
\begin{eqnarray}
D(U;\nu_a)=\mbox{det}\left(
\begin{array}{cccccc}
\mbox{id}& -R_1A^{-1}	 & 0	  & \cdots     	& 0 & | 0 \\
\hline\\
0	&\mbox{id}&        &     	&     & | 0 \\
0 	& 	 &\mbox{id} &     	&     & | 0 \\
\cdots 	&        &        & \cdots    	&     &     \\
0 	&        &        & 		&\mbox{id}& | 0 \\
\hline\\
0	& 0      & \cdots &             & 0   & |\mbox{id}
\end{array}
\right)
[\frac{\partial\bar{F}^{aA}\nu_a}{\partial U^B}]
\end{eqnarray}
\begin{eqnarray}
=\mbox{det}\left(
\begin{array}{cccccc}
K-RA^{-1}S& 0    & R_2	  &\cdots      	& R_N & | * \\
\hline\\
0 	& A      &        &     	&     & | 0 \\
0 	& 	 & A      &     	&     & | 0 \\
\cdots 	&        &        & \cdots    	&     &     \\
0 	&        &        & 		& A   & | 0 \\
\hline\\
\mbox{*}& 0      & \cdots &             & 0   & | *
\end{array}
\right),
\end{eqnarray}
we may continue in this manner
until all $R_\lambda$ have been removed:
\begin{eqnarray}
D(U;\nu_a)=\mbox{det}\left(
\begin{array}{cccccc}
K-RA^{-1}S& 0    & \cdots &           	& 0   & | * \\
\hline\\
0 	& A      &        &     	&     & | 0 \\
0 	& 	 & A      &     	&     & | 0 \\
\cdots 	&        &        & \cdots    	&     &     \\
0 	&        &        & 		& A   & | 0 \\
\hline\\
\mbox{*}& 0      & \cdots &             & 0   & | *
\end{array}
\right)
\end{eqnarray}
It follows that
\begin{eqnarray}
D=|A|^N\mbox{det}\left(
\begin{array}{cc}
K-RA^{-1}S& | * \\
\hline\\
\mbox{*}& 	| *
\end{array}
\right)\equiv|A|^N\mbox{det}[D^B_A](U;\nu_a).
\end{eqnarray}

\underline{Step (c):}
It may be
readily verified that
\begin{eqnarray}
\begin{array}{ll}
A^{-1}=&-(u^c\nu_c)^{-1}
[g^b_a
+u^bu_a
-\nu^{-2}\nu^b\nu_a],\\
\mbox{det}A=&\nu^2(u^c\nu_c)^2.
\end{array}
\label{A1}
\end{eqnarray}
Multiplication of $[D^B_A](U;\nu_a)$ by an appropriate
matrix with unit determinant from the $right$, whereby multipling
the leading column block by $A$ from the right,
leads us to consider
\begin{eqnarray}
\begin{array}{ll}
D(U;\nu_a)&=|A|^{N-1}\mbox{det}\left(
\begin{array}{cc}
(KA-RA^{-1}SA& | * \\
\hline\\
\mbox{*}A& 	| *
\end{array}
\right)\\
&\equiv |A|^{N-1}\mbox{det}[\bar{D}^B_A](U;\nu_a).
\end{array}
\end{eqnarray}

\underline{Step (d):} It remains to ascertain that
$\mbox{det}[\bar{D}^B_A](U;\nu_a)$ constitutes a polynomial
in $\nu_a$. This is not immediate, in view of the presence
of $A^{-1}$ in $[\bar{D}^B_A]$ and the factor
$(u^c\nu_c)^{-1}$ in (\ref{A1}).
As it turns out, any rational factor thus introduced
in $\mbox{det}[\bar{D}^B_A](U;\nu_a)$ will fortuitously be cancelled.
This will be shown below.

Explicit evaluation of $A^{-1}S^\lambda A$ gives
\begin{eqnarray}
A^{-1}S^\lambda A=(u^c\nu_c)^{-1}[(\nu^2h^{\lambda b}
-(h^{\lambda c}\nu_c)\nu^b)u_a]+Q^\lambda,
\end{eqnarray}
where
$Q^\lambda=(h^{\lambda c}\nu_c)\mbox{id}
+[u^b\{(u^c\nu_c)h^{\lambda}_a-(h^{\lambda c}\nu_c)u_a]$
represents the remaining matrix whose entries form
polynomial expressions in $\nu_a$.
This gives
\begin{eqnarray}
RA^{-1}S A=\nu^2(u^c\nu_c)^{-1}
[(h^2\nu^b-\tilde{h}^b)u_a]+RQ,
\label{sing}
\end{eqnarray}
where $RQ\equiv\sum_\lambda R_\lambda Q^\lambda$
and $\tilde{h}^b=\sum_\lambda(h^c_\lambda\nu_c)h^{\lambda b}$.
This shows that $(u^c\nu_c)^{-1}$ enters in
a rank-one update to the ($\nu_a$- polynomial) $KA-RQ$, and, therefore,
introduces a factor $(u^c\nu_c)^k$ with $k\ge-1$
in each term in
$\mbox{det}[\bar{D}^B_A](U;\nu_a)$. This follows
by Lemma 4.1.
More elementary,
we may introduce a special frame of reference in which
$u_a=(-1,0,0,0)$. $(u^c\nu_c)^{-1}$ then simply enters in the first
column
of $[\bar{D}^B_A](U;\nu_a)$,
and the forementioned observation
readily follows.

Inspection of the full
matrix $[\bar{D}^B_A](U;\nu_a)$ with $U=(u^b,h^{\lambda b},P,r)$,
\begin{eqnarray}
[\bar{D}^B_A](U;\nu_a)=
\left(
\begin{array}{ccc}
KA-RA^{-1}SA& | q_P(u^c\nu_c)[u^b] + [\nu^b]
& |q_r(u^c\nu_c)[u^b]\\
\hline\\
r[\nu_cA^c_a]		 & | 	0   & | (u^c\nu_c)\\
(\xi^c\nu_c)[u_cA^c_a]	 & | 	0   & | 0
\end{array}
\right),
\end{eqnarray}
where $q_y=\partial q/\partial y$,
$q=rf+h^2$, further shows that the last column
is proportional to $(u^c\nu_c)$.
This dependency introduces a factor
$(u^c\nu_c)$ in $\mbox{det}[\bar{D}^B_A](U;\nu_a)$ which,
combined with the foregoing, forces
each term in
$\mbox{det}[\bar{D}^B_A](U;\nu_a)$ to contain
$(u^c\nu_c)^k$ with $k\ge0$:
$\mbox{det}[\bar{D}^B_A](U;\nu_a)$ forms a polynomial expression
in $\nu_a$ (homogeneous of degree 10).

Together, the results from Step (c) and Step (d) show that
$|A|^{N-1}$ is a factor in the
the $6+4N$-degree homogeneous polynomial $D(U;\nu_a)$. $\Box$
\mbox{}\\
\mbox{}\\
The full factorization is now contained in
\newtheorem{thcd}{Theorem}[section]
\begin{thcd}
Yang-Mills' characteristic determinant is of the form
\begin{eqnarray}
D(U;\nu_a)=-(\xi^c\nu_c)(u^c\nu_c)^{2N-1}(\nu^c\nu_c)^{N}Y(U;\nu_a),
\label{factD}
\end{eqnarray}
where $Y(U;\nu_a)$ is a sixth-order polynomial, homogeneous in $\nu_a$.
\end{thcd}

$Proof:$
Inspection of $[\bar{D}^B_A](U;\nu_a)$ as defined in the proof of Lemma
4.1
shows that all its entries are even in $u^b$ except for the lower row
\begin{eqnarray}
[r[\nu_cA^c],0,(u^c\nu_c)]
\end{eqnarray}
stemming from conservation of baryon number, $\nabla_a(ru^a)=0$.
$\mbox{det }[\bar{D}^B_A](U;\nu_a)$ is therefore odd in $u^b$.
In view of $|A|=(\nu^c\nu_c)(u^c\nu_c)^2$, this forces
$D(U;\nu_a)$ to be odd in $u^b$.\footnote{In \cite{mvp:91} the
characteristic determinant of ideal MHD ($N=1$)
is even in $u^b$, as we work there with
$u^a\nabla_a S=0$ (instead of $\nabla_a\{(u^2+1)\xi^a\}=0$).}
Now notice that $u^b$ appears only through $(u^c\nu_c)$ in
the scalar field $D(U;\nu_a)$,
since
$u^ch^\lambda_c=0$ for all $\lambda$ and $u^cu_c=-1$
(see also Proposition 5.1 in \cite{mvp:91}).
The roots of
$D(U;\nu_a)$ are invariant quantities,
necessarily invariant under
a sign-change $u^b\rightarrow-u^b$. It follows that
an odd factor of $(u^c\nu_c)$ must factor in
$D(U;\nu_a)$, leaving all other factors even
in $(u^c\nu_c)$.
The factor $(u^c\nu_c)^{2N-2}$ introduced by
$|A|^{N-1}$, therefore, forces
$(u^c\nu_c)^{2N-1}$ to be a factor in
$\mbox{det }D(U;\nu_a)$.

The factor $(\nu^c\nu_c)^N=\nu^{2N}$ in (\ref{factD}), rather than
$\nu^{2N-2}$ as promised by $|A|^{N-1}$, follows from
the fact that the $6\times4N$ dimensional elements
\begin{eqnarray}
[0_a,\mu_1\nu_a,\cdots,\mu_N\nu_a,0,0],
\end{eqnarray}
$\mu_k\epsilon{\cal C}$,
form left
null elements of the characteristic matrix $[F^B_A](U:\nu_a)$,
whenever $\nu^2=0$. This
follows from $\nu_a$ being
a null element of $both$ $A^b_a$ and $S^{\lambda b}_a$, $i.e.$,
\begin{eqnarray}
\nu_cA^c_a=\nu_cS^{\lambda c}_a=0,
\end{eqnarray}
whenever $\nu^2=0$. This completes the proof.$\Box$
\mbox{}\\
\mbox{}\\

We conclude that the null cone ${\cal N}$ at each point on the
space-time manifold is described by
\begin{eqnarray}
\begin{array}{ll}
\mbox{(1) Entropy waves: }   & u^c\nu_c=0,\\
\mbox{(2) Yang-Mills waves: }& Y(U;\nu_a)=0.
\end{array}
\end{eqnarray}
Thus, the Yang-Mills waves are completely defined by
a sixth-order polynomial $Y(U;\nu_a)$.

Recall that in the theory of ideal magneto-hydrodynamics
(MHD) we find the hydrodynamical waves
and the Alfven waves through the zeros
of a sixth order polynomial, which factorizes over
a fourth-order (hydrodynamical waves) and
a second-order (Alfven waves) polynomial \cite{cb:a,mvp:91}.
However, one may not expect
such factorization to persist in the more
general case of a Yang-Mills fluid.

Expressions (27)-(28) show that the
Yang-Mills magnetic stress-tensor $H^b_a$, as defined above,
represents the entire magnetic field $h_a\epsilon{\cal G}$
in the infinitesimal wave structure. It follows that
the Yang-Mills waves $Y(U;\nu_a)=0$ are completely
determined by $(u^a,H^b_a,P,r)$. Because $H_{ab}$ is
real-symmetric, the Spectral Decomposition Theorem
allows for the representation
\begin{eqnarray}
H_{ab}=\sum_{i=1}^3 s_a^{(i)}s_b^{(i)},
\end{eqnarray}
the $s^{(i)}_a$, $i=1,2,3$, being three one-forms on the
space-time manifold.
Since $H_{ab}u^a\equiv0$, it follows that
$N=3$ without loss of generality in the structure
of the infinitesimally small amplitude waves in
ideal Yang-mills fluids.

The new wave features in Yang-Mills fluids in ideal
Yang-Mills fluids are contained in the
normal cone as defined by the
zeros of the {\em Yang-Mills characteristic polynomial} $Y(U;\nu_a)$.
Of foremost importance
is to establish hyperbolicity, $i.e.$, to establish that
$Y(U;\nu_a)=0$ contains a full set of three (real-valued) normal
conical sheats which are everywhere space-like. Of course, we need
not require strict, algebraic hyperbolicity, that is, some of these normal
sheats may (partially) coincide, associated with
multiplicities of the roots of
$Y(U;\nu_a)=0$. This may be studied through
symmetrization.

\section{On hyperbolicity of systems of conservation laws}

Friedrichs \& Lax \cite{fr:71}
and Friedrichs \cite{fr:74} conceived the deep
and general result that a system of conservation laws is well-posed
when the system (implicitly) contains a conservation law for
a convex quantity. A quantity of this kind naturally enforces an integral
bound on the solution for all time.
More specific regularity results for finite time are obtained
following symmetrization, as obtained by such conserved convex quantity.
The resulting system is in {\em symmetric hyperbolic} form
(\cite{fr:71}), which establishes at one stroke both hyperbolicity
and well-posedness.
In our study of the Yang-Mills equations, a slightly weaker version
of Friedrichs' symmetrization procedure is formulated in view
of the conserved, algebraic constraints
in the theory of ideal Yang-Mills fluids.

\subsection{Friedrichs' symmetrization procedure}

Ideal Yang-Mills as given in Theorem 3.1
is a particular case of a system of $M$ conservation laws of the form
\begin{eqnarray}
\nabla_a F^{aB}=f^B.
\label{FL}
\end{eqnarray}
Friedrichs \cite{fr:74}
introduced certain Properties CI and CII, discussed below,
which are sufficient
for (\ref{FL}) to constitute
a {\em symmetric hyperbolic} system,
\begin{eqnarray}
A^{aAB}(V)\nabla_a V_A=
A^{tAB}(V)\nabla_t V+A^{\alpha AB}(V)\nabla_\alpha V_A=f^B(V),
\label{SYA}
\end{eqnarray}
characterized by symmetry of the $A^{aAB}$ (in $A$ and $B$)
and positive definiteness of $A^{tAB}$.
Systems of this type are known to give rise to
well-posed initial value
with sensitivity on initial data of finite order.
In the context of general relativity, the
requirement that $A^{aAB}\xi_a$ is nonsingular for all
time-like one-forms $\xi_a$ furthermore ensures that
no infinitesimally small amplitude wave exceeds
unit velocity.

Friedrichs' \cite{fr:74} symmetrization procedure
applies to systems possessing what Friedrichs calls
a ``main dependency" relation of the form
\begin{eqnarray}
\begin{array}{lc}
\mbox{CI}:&W_A\delta F^{aA}\equiv0.
\end{array}
\label{PCI}
\end{eqnarray}
Here, $W_A$ is a nontrivial vector
and $\delta$ denotes a total variation (this property is
called $\mbox{CI}^\prime$ in [Friedrichs, 1974]).
Evidently, this requires $W_Af^A\equiv0$.
Then differentiation with respect to any $V^A$ yields
\begin{eqnarray}
\frac{\partial W_A}{\partial V^C}
\frac{\partial F^{aA}}{\partial V^D}\nabla_a V^D
+W_A\frac{\partial^2 F^{aA}}{\partial V^C\partial V^D}
\nabla_a V^D
=0.
\label{symm0}
\end{eqnarray}
This shows that for each $a$
\begin{eqnarray}
\frac{\partial W_A}{\partial V^C}
\frac{\partial F^{aA}}{\partial V^D}
\label{symm}
\end{eqnarray}
is symmetric (in $C$ and $D$). If, furthermore, $V^A$ is such that
for some time-like $\xi_a$
\begin{eqnarray}
\begin{array}{lc}
\mbox{CII:}&
\delta W_A
\delta F^{aA}\xi_a
=\delta V^C(\frac{\partial W_A}{\partial V^C}
\frac{\partial F^{aA}\xi_a}{\partial V^D})\delta V^D
\end{array}
\end{eqnarray}
is positive definite
(for all $\delta V^A$),
then (\ref{FL}) naturally
constitutes a symmetric hyperbolic
system in $V_A$ in the form of
\begin{eqnarray}
\frac{\partial W_A}{\partial V^C}
\frac{\partial F^{aA}}{\partial V^D}
\nabla_a V^D=
\frac{\partial W_A}{\partial V^C}f^A.
\label{FFF}
\end{eqnarray}
Writing $Z^A=\nabla_aF^{aA}-f^A$, it follows that solutions
to (\ref{FFF}) satisfy
\begin{eqnarray}
\left\{
\begin{array}{l}
\frac{\partial W_A}{\partial V^C}Z^A=0,\\
W_AZ^A=0,
\label{p2}
\end{array}
\right.
\end{eqnarray}
where the second equation is a consequence of Property CI.
If Property CII is satisfied as well,
$(W_A,\frac{\partial W_A}{\partial V^C})$ can be readily
seen to be nonsingular, so that
any solution
to (\ref{p2}), is, in fact, a solution to (\ref{FL}).

This summarizes the symmetrization procedure
constructed by Friedrichs \cite{fr:74} for systems possessing
a single dependency relation (\ref{PCI}).
It should be mentioned that Friedrichs \cite{fr:74} also
treats the case of more than one dependency relations.
For the purpose
of symmetrization of ideal Yang-Mills in divergence form
there exists precisely one, main dependency relation, although
in a slightly weaker form than CI as stated above. Not
suprisingly, this is due to the fact that the divergence
formulation allows for a larger class of solutions, which
includes
the nonphysical solutions with
$c_\lambda\ne0$.

\subsection{Symmetrization in the presence of conserved constraints}

Ideal Yang-Mills in divergence form contains
the constraints $c_\lambda=0$ as conserved quantities when
initial data satisfy certain compatibility conditions.
The system itself allows for solutions with $c_\lambda\ne0$ as well,
in response to nonphysical initial data.
It is therefore not suprising that in this generality ideal
Yang-Mills in divergence form does not satisfy conditions
CI or CII.
Rather, we require a main dependency relation of the form
\begin{eqnarray}
\begin{array}{lc}
\mbox{YI}:&W_A\delta F^{aA}\equiv\delta z^a
\end{array}
\label{PYI}
\end{eqnarray}
for some nontrivial $W_A$ and some vector field $z^a$. We will
later require that $z^a$ vanishes for solutions which satisfy
the constraints.
Clearly, Property YI is weaker than
Friedrichs' Property CI.
However, the identity
\begin{eqnarray}
\frac{\partial W_A}{\partial V^C}
\frac{\partial F^{aA}}{\partial V^D}\nabla_a V^D
+W_A\frac{\partial^2 F^{aA}}{\partial V^C\partial V^D}\nabla_a V^D
=\frac{\partial^2 z^a}{\partial V^C\partial V^D}\nabla_a V^D,
\label{symm1}
\end{eqnarray}
instead of (\ref{symm0}),
clearly does not change the symmetry of the coefficient matrices
(\ref{symm}). In what follows,
nonsingularity of
$(W_A,\frac{\partial W_A}{\partial V^C})$ will be assumed,
so that (\ref{p2}) enforces equivalence of (\ref{symm1}) with
(\ref{FL}).

For solutions with $c_\lambda\ne0$, different from
the Yang-Mills solutions $c_\lambda=0$, we may not expect
convexity property CII to hold.
We do, however, wish solutions to ideal Yang-Mills
satisfying the constraints
\begin{eqnarray}
c_0=0,\\
c_\lambda=0,
\end{eqnarray}
to satisfy a symmetric hyperbolic system. To this end, a
slightly weaker formulation of CII will be used.

\newtheorem{def1}{Definition}[section]
\begin{def1}
Consider a scalar $Q=Q(U^B,\delta U^C)$
and a set of scalar constraints, $c_k(U^C)=0$.
The variation $\delta U^C\ne0$ satisfying $\delta c_k(U)=0$ will
be referred to as the {\bf constraint variation} of $U^C$
with respect to the $c_k=0$.
If $Q>0$ with respect to the constraint variation of $U^C$,
then $Q$ is said
to be {\bf constraint positive definite} with respect to the $c_k=0$.
\end{def1}
Notice that for each $U^B$ the constraints in Definition 5.1 define
a linear subspace of variations $\delta U^C$.

We now define
constraint positive definiteness of
\begin{eqnarray}
\begin{array}{lc}
\mbox{YII:}&Q:=
\delta W_A
\delta F^{aA}\xi_a
=\delta V^C(\frac{\partial W_A}{\partial V^C}
\frac{\partial F^{aA}\xi_a}{\partial V^D})\delta V^D
\end{array}
\end{eqnarray}
for some time-like $\xi_a$
as Property YII. Clearly, YII is
weaker than Friedrichs' Property CII.

Property YI and YII are sufficient for (\ref{FL}) to imply
a symmetric hyperbolic
system in $V_A$.
To see this, we use
the following construction.

\newtheorem{lm1}{Lemma}[section]
\begin{lm1}
Given a real-symmetic
$A\epsilon{\cal L}({\bf R}^n,{\bf R}^n)$
which is positive definite on a linear subspace
${\cal V}\subset{\bf R}^n$, there exists
a real-symmetric, positive definite
$A^*\epsilon{\cal L}({\bf R}^n,{\bf R}^n)$ such that
\begin{eqnarray}
A^*y=Ay\mbox{    }(y\epsilon{\cal V}).
\end{eqnarray}
\end{lm1}

$Proof:$  Let ${\cal V}^\perp$ denote the orthogonal
complement of ${\cal V}$. If ${\cal V}^\perp=\{0\}$,
we may take $A^*=A$. Otherwise, pick
$x\epsilon {\cal V}^\perp$ with $||x||=1$, and form
\begin{eqnarray}
A_1=A+\mu xx^T,
\end{eqnarray}
where $x^T$ denotes the transpose of $x$. $A_1$ is
manisfestly symmetric. We will choose
$\mu$ so that $A_1$ is positive definite
on the space ${\cal V}_1$ of all vectors of the form
$z=y+\lambda x$, where $y\epsilon{\cal V}$ and
$\lambda$ is a real scalar.

To this end, evaluate
\begin{eqnarray}
\begin{array}{ll}
z^TAz&=y^TAy+2\lambda x^TAy+\lambda^2x^TAx+\lambda^2\mu^2\\
&\ge c||y||^2-2|\lambda|M||y||+\lambda^2(\mu^2-M)\\
&=(\mu^2-M)(|\lambda|-\frac{M||y||}{\mu^2-M})^2+c||y||^2-
\frac{M^2}{\mu^2-M}||y||^2.
\label{est}
\end{array}
\end{eqnarray}
Here, $c>0$ is a coercivity constant for $A$ on ${\cal V}$, $i.e.$,
$y^TAy\ge c||y||$ ($y\epsilon{\cal V}$), and $M=||A||$.
For sufficiently large $\mu>M^{1/2}$, we have
\begin{eqnarray}
c||y||^2
-\frac{M^2}{\mu^2-M}||y||^2\ge\frac{c}{2}||y||^2.
\end{eqnarray}
Writing $\lambda=a||y||$, (\ref{est}) yields
\begin{eqnarray}
z^TAz\ge K(a)||z||^2,
\end{eqnarray}
where
$K(a)=(1+a^2)^{-1}
\{
(\mu^2-M)(a-\frac{M}{\mu^2-M})^2+\frac{c}{2}
\}$.
Clearly,
$\mbox{inf}_{a\epsilon{\bf R}}
K(a)\equiv c_1>0$.

We have now proved that there exists an embedding of
$A|_{\cal V}$ in a symmetric $A_1$ which is positive definite
on ${\cal V}_1$.

Repeating the construction above
$k=\mbox{dim}{\cal V}^\perp$ times, we may exhaust ${\cal V}^\perp$,
thereby arriving at a symmetric $A^*=A_k$ which is
positive definite on ${\cal V}_k={\bf R}^n$.
This completes the proof. $\Box$\\
\mbox{}\\
\mbox{}\\
Consider a system (\ref{FL}) which satisfies
Property YI, thereby arriving at a system (\ref{SYA}),
which may, alternatively, be written as
\begin{eqnarray}
A^{tAB}(V)\partial_t V+A^{\alpha AB}(V)\partial_\alpha V_A=
\bar{f}^B(V).
\label{shyp0}
\end{eqnarray}
Here, $\bar{f}^B$ now contains the contributions from the
connection symbols as well.
Assuming that (\ref{FL}) satisfies Property YII
with $\xi^b=(\partial_t)^b$,
we may form a positive definite $(A^{tAB})^*$ following Lemma 5.1.
Indeed,
for $V$ such that $c_\lambda(V)=0$ for all $\lambda$,
Property YII implies $A^{tAB}(V)$ is positive definite on
${\cal V}=\bigcap_\lambda
\{\delta V|\frac{\partial c_\lambda}{\partial V^C}\delta V_C=0\}$.
According to Lemma 5.1, we may insist
\begin{eqnarray}
(A^{tAB})^*y=A^{tAB}y
\end{eqnarray}
for all constraint variations $y\epsilon{\cal V}$.

We are thus led to consider
the symmetric hyperbolic system
\begin{eqnarray}
(A^{tAB})^*(V)\partial_t V+A^{\alpha AB}(V)\partial_\alpha V_A
=\bar{f}^B(V).
\label{shyp}
\end{eqnarray}

The arguments above show that solutions to the original
divergence system of equations (\ref{FL}) in response to initial data
compatible with (\ref{COMP}) (which have the property that
$c\equiv0$ throughout)
are solutions to the symmetric hyperbolic system of equations (\ref{shyp}).
It naturally follows that these solutions
depend continuously on their data as described in \cite{fm:72}.

Friedrichs's symmetrizations procedure and
the symmetrization procedure in the presence of conserved
constraints as presented in this Section are
summarized in Table I.
\begin{table}
\begin{tabular}{|c|c|}\hline\hline
{\em Friedrichs' procedure}
&{\em In this paper, with conserved $c=0$}\\ \hline
\multicolumn{1}{|c|}{\em Properties Cx}&
\multicolumn{1}{c|}{\em Properties Yx}\\ \hline
CI: $W_A\delta F^{aA}=0$ &YI: $W_A\delta F^{aA}=\delta z^a$\\
CII: $\delta W_A\delta F^{aA}\xi_a$
&YII: $\delta W_A\delta F^{aA}\xi_a$\\
positive definite & constraint positive definite\\
(CIII when more than one&
YIII: $c\equiv0$ if (\ref{COMP}) satisfied\\
dependency relation \cite{fr:74}) &\\
\hline\mbox{}&\\
$\frac{\partial W_A}{\partial V^C}
\frac{\partial F^{aA}}{\partial V^D}
\nabla_a V^D=
\frac{\partial W_A}{\partial V^C}f^A$&
$(\frac{\partial W_A}{\partial V^C}
\frac{\partial F^{tA}}{\partial V^D})^*\partial_t V^D
+\frac{\partial W_A}{\partial V^C}
\frac{\partial F^{\alpha A}}{\partial V^D}\partial_\alpha V^D
=\frac{\partial W_A}{\partial V^C}
\bar{f}^B$\\
\mbox{}&\\
\hline\hline
\end{tabular}
\caption{Symmetrization procedures: summary of
Friedrichs' symmetrization procedure and the
symmetrization procedure in the presence of
conserved constraints, as presented in this paper.
Friedrichs' procedure requires a system $\nabla_aF^{aB}=0$
to possess Properties CI and CII (in the presence of a
single dependency relation), and the procedure presented
here requires the system to possess Properties YI, YII
and YIII.}
\end{table}

It would be of interest to also use the symmetric hyperbolic system of
equations
(\ref{shyp})
for existence proofs of solutions to the divergence
formulation of ideal Yang-Mills fluids. To this end,
solutions to initial
value problems for the symmetric hyperbolic system of equations
(\ref{shyp})
in response to data compatible with
(\ref{COMP}) must be shown to have the property that
$c\equiv0$ throughout.
However, this falls outside the scope of this work.

\section{Ideal Yang-Mills in symmetric hyperbolic form}

We will show that ideal Yang-Mills
satisfies a symmetric hyperbolic system of the form (\ref{SYA}).
Using the symmetrization procedure in the presence of
constraints as outlined in the previous section, this result
follows by showing that ideal Yang-Mills in divergence form
(\ref{PDE}) satisfies Properties YI and YII. The proof of Theorem 3.1
establishes that (\ref{PDE}) satisfies Property YIII.
It is well-known that the equations of relativistic
hydrodynamics
can be written in symmetric hyperbolic form
\cite{an:89,rs:81}.
The desired result, therefore, will be established
``bootstrap-wise," by showing that
ideal Yang-Mills can be written in symmetric hyperbolic form,
whenever the equations of relativistic hydrodynamics
can be written in symmetric hyperbolic form.

For the purpose of symmetrization, we use a second
divergence form of ideal Yang-Mills fluids (following
\cite{mvp:91}), in which the equations of hydrodynamics
are more similar to those already discussed in the literature:
\begin{eqnarray}
\hat{\nabla}_aF^{aA} \equiv
\left\{
\begin{array}{c}
\nabla_a T^{ab} = 0,\\
\hat{\nabla}_a \{h^{[a}u^{b]}+g^{ab}c(U)\} = 0,\\
\nabla_a (ru^a) = 0,\\
\nabla_a (rSu^a)=0.
\label{PDE2}
\end{array}
\right.
\end{eqnarray}
This system may readily be seen to be equivalent to (\ref{PDE})
for continously differentiable solutions
in view of the thermodynamic relation
$\mbox{d}P=r\mbox{d}f-rT\mbox{d}S$.
Note, however, that the jump conditions across surfaces
of discontinuity as follow from
a weak formulation for this system (\ref{PDE2})
are improper (see \cite{mvp:93b} for a discussion
on this point in magneto-hydrodynamics).
We organise (\ref{PDE2}) as follows
\begin{eqnarray}
\nabla_a F^{aB}_f
+\nabla_a F^{aB}_m
=f^B_f+f^B_m
\end{eqnarray}
where
\begin{eqnarray}
\begin{array}{ll}
\nabla_a F^{aB}_f\equiv
\left\{
\begin{array}{l}
\nabla_a T^{ab}_f,\\
\nabla_a (ru^a),\\
\nabla_a \{\xi^a(u^2+1)\},
\end{array}
\right.
  &f_f^B=0,\\
\mbox{}\\
\nabla_a(F^{aA}_m)_\lambda \equiv
\left\{
\begin{array}{l}
\nabla_a T^{ab}_m,\\
\nabla_a \omega^{ab}_\lambda,
\end{array}
\right.
  &f_m^B=\left\{
  \begin{array}{l}
  0,\\
  -c_{\lambda\mu\nu}A^\mu_e\omega^{\nu eb}.
  \end{array}
  \right.\\
\end{array}
\end{eqnarray}
Here, $c_{\lambda\mu\nu}$ are the structure constants
of the Yang-Mills Lie algebra, ${\cal G}$.
As mentioned before, these equations must be closed by
(\ref{closure}),
\begin{eqnarray}
\nabla_aA_b-\nabla_bA_a
=F_{ab}
-[A_a,A_b]\equiv G_{ab}.
\label{gauge0}
\end{eqnarray}
\mbox{}\\
\mbox{}\\
Before turning to the symmetrization of the nonlinear
system of equations (\ref{PDE2}), we wish to remark
that with temporal gauge
the closure equations can be seen explicitly to pose no difficulties in
the symmetrization process.
To see this, let $\xi^b$ is any given time-like vector field, and
consider the equations
\begin{eqnarray}
(\xi^c\nabla_c)A_a-\xi^c\nabla_aA_c
=\xi^cG_{ca}.
\label{gauge}
\end{eqnarray}
Because $\hat{\nabla}_a\omega^{ab}=0$ enforces
$\hat{\nabla}_{[a}F_{bc]}
=\nabla_{[a}F_{bc]}+[A_{[a},F_{bc]}]=0$,
and $[F_{[ab},A_{c]}]\equiv
\nabla_{[a}[A_b,A_{c]}]$, the right hand-side in
(\ref{gauge0}) satisfies
$\nabla_{[a}G_{bc]}=0$ (using the result that $c(U)=0$).
Denoting with Greek subscripts contractions with
$\xi^b-$orthogonal vectors, (\ref{gauge}) thus
may be seen to ensure conservation of
$\nabla_\alpha A_\beta-\nabla_\beta A_\alpha=G_{\alpha\beta}$.
Eqs. (11)-(12) therefore impose (75) as evolution equations for
$A_a$.
With conservation $\xi^cA_c=0$, any solution to (\ref{gauge})
satisfies
\begin{eqnarray}
(\xi^c\partial_c)A_a=\xi^cG_{ca}+A_c\partial_a\xi^c
\end{eqnarray}
with coefficient matrices
\begin{eqnarray}
(\xi^c\nu_c)g^b_a.
\end{eqnarray}
For this reason,
the closure equations (\ref{closure}) with temporal gauge
need no further consideration
in the symmetrization of the equations of
ideal Yang-Mills in divergence form.
\mbox{}\\
\mbox{}\\
In the variables $(u^b,h^b,f,S)$ (\ref{PDE2}) is, of course,
regular in the sense of Cauchy-Kowalewski (by its equivalence
to (\ref{PDE})), and possesses therefore no dependency relation.
If we consider (\ref{PDE2}) in
$V=(v_\alpha,h^b,f,S)$, however, where $v_\alpha$ is
a three-parametrization of $u^b$ such that $u^2\equiv-1$,
this system becomes overdetermined by one and a main
dependency relation results (as in Property YI).
This will be made explicit
below.

\subsection{Symmetrization of hydrodynamics}
The equations of relativistic hydrodynamics,
\begin{eqnarray}
\nabla_a F^{aB}_f=0,
\end{eqnarray}
as follow from ideal Yang-Mills in divergence form with
vanishing Yang-Mills field, may be written in symmetric
hyperbolic form following the arguments given by
Friedrichs \cite{fr:74}, Ruggeri \& Strumia \cite{rs:81} and
Annile \cite{an:89}.
Friedrichs \cite{fr:77} emphasis that symmetric hyperbolic
systems will not generally be in covariant form;
the construction of
systems of this type is intended as a method
of establishing well-posedness, an aspect of which $is$
frame independent.
For symmetrization of the equations of relativistic hydrodynamics,
the velocity four-vector, $u^b$, may be represented by
a three-parameter representation, $u^b=u^b(v_\alpha)$, in view
of $c_0=u^cu_c+1=0$.
Consider $W^f_A=\frac{1}{T}(u_a,f-TS,T)$ as a function of
$V^f_C=(v_\alpha,T,f)$.
Notice that variations $\delta V^f_A$ satisfy
$\delta u^2=0$ by construction.
In this parametrization, both Friedrichs' Properties CI \& CII
hold true. Indeed,
we have a main dependency relation and an
associated quadratic, $Q_f$, of the form \cite{rs:81}
\begin{eqnarray}
\begin{array}{lll}
W^f_A\delta F^{aA}_f&=&u_b\delta T^{ab}_f+(f-TS)\delta(ru^a)+
T\delta (rSu^a)\equiv0,\\
Q_f:=\delta W_A\delta F^{aA}_f
&=&T\{\delta[\frac{u_b}{T}]\delta T^{ab}_f
+\delta[\frac{f-TS}{T}]\delta(ru^a)\}
+\frac{\delta T}{T}
W^f_A\delta F^{aA}_f\\
&\equiv&
T\{\delta[\frac{u^b}{T}]\delta T^{ab}_f+
\delta[\frac{f-TS}{T}]\delta(ru^a)\}.
\end{array}
\label{HC}
\end{eqnarray}
Ruggeri \& Strumia \cite{rs:81} and
Anile \cite{an:89} have shown $Q_f$ to be positive definite,
provided that the following conditions on the
free enthalpy, $G$, and the sound velocity, $a$,
are satisfied \cite{an:89}:
\begin{eqnarray}\left\{
\begin{array}{ll}
\mbox{  the free enthalpy:  }&-G(T,P)\mbox{  is convex},\\
\mbox{  the sound velocity:  }&a^{-2}=\frac{f}{r}
\frac{\partial r}{\partial f}|_S>1.
\end{array}
\label{H2}
\right.
\end{eqnarray}
We will now proceed with the full equations of ideal Yang-Mills
in divergence form.

\subsection{Symmetrization with Yang-Mills field}
Ideal Yang-Mills in divergence form will be shown to be
satisfying Property YI and YII, provided that the hydrodynamical
conditions (\ref{H2}) have been met.

{\em Property YI.} Recall from (5) and (14)
the contribution to the stress-energy tensor by the magnetic field,
$T^{ab}_m$, and the new Yang-Mills tensor, $\omega^{ab}$, respectively, given
by
\begin{eqnarray}
T^{ab}_m=h^2u^au^b+\frac{h^2}{2}g^{ab}-h^ah^b,\\
\omega^{ab}=h^au^b-u^ah^b+g^{ab}h^cu_c.
\end{eqnarray}
Let us consider $W_A=\frac{1}{T}(u_a,h_a,f)$ with
parametrization $V_A=(v_\alpha,h_a,T,f)$.
The total variation, $\delta$, then satisfies $c_0(U):=u^2+1\equiv0$
by construction, as before.
The variational expressions
\begin{eqnarray}
\begin{array}{lll}
u_b\delta T^{ab}_m&=&u_b\{h^2u^a\delta u^b+h^2u^b\delta u^a
+2u^au^bh_c\delta h^c\\
&&+g^{ab}h_c\delta h^c-h^b\delta h^a
-h^a\delta h^b\}\\
&=&-h^2\delta u^a-u^a(h_c\delta h^c)-h^a(u_c\delta h^c)-c\delta h^a,\\
h_b\delta\omega^{ab}&=&h_b\{h^a\delta u^b+u^b\delta h^a
-h^b\delta u^a-u^a\delta h^b+g^{ab}\delta c\}\\
&=&
h^a(h_c\delta u^c)
+c\delta h^a
-h^2\delta u^a
-u^a(h_c\delta h^c)
+h^a\delta c,
\end{array}
\end{eqnarray}
introduce the identity
\begin{eqnarray}
u_b\delta T^{ab}
-h_b\delta\omega^{ab}
=-2\delta(ch^a)\equiv\delta z^a.
\end{eqnarray}
This establishes that Property YI is satisfied:
\begin{eqnarray}
W_A\delta (F^{aA}_f+F^{aA}_m)
=\delta z^a.
\end{eqnarray}

{\em Property YII.} Ideal Yang-Mills in divergence form
can be shown to be satisfying Property YII by
consideration of the extension to
relativistic hydrodynamics as introduced by
the Yang-Mills fields.

\newtheorem{cv1}{Proposition}[section]
\begin{cv1}
If the equations of relativistic hydrodynamics
are convex, then ideal Yang-Mills in divergence form
is constraint convex.
\end{cv1}

$Proof:$
Let $Q_f$ denote the quadratic variation associated with
the equations of relativistic hydrodynamics as in (\ref{HC}),
and suppose conditions (\ref{H2}) have been met.

Consider the variation, $\delta$, introduced by the
total variation of $V_C$ ($i.e.$, variations satisfying
$\delta c_0(U)=\delta (u^2+1)\equiv0$).
The variational expressions
\begin{eqnarray}
\begin{array}{lll}
\delta u_b\delta T^{ab}_m\xi_a&=&
h^2(u^c\xi_c)(\delta u_c\delta u^c)^2\\
& &+(\xi_c\delta u^c)(h_c\delta h^c)
-(h_c\delta u^c)
(\xi_c\delta h^c)
-(\xi_c h^c)(\delta u_c\delta h^c),\\
\delta h_b\delta\omega^{ab}\xi_a&=&
(h^c\xi_c)(\delta h_c\delta u^c)+(\xi_c\delta h^c)(u_c\delta h^c)\\
& &-(\xi_c\delta u^c)(h_c\delta h^c)
-(u^c\xi_c)(\delta h_c\delta h^c)+(\xi_c\delta h^c)\delta c,
\end{array}
\end{eqnarray}
introduce the identities
\begin{eqnarray}
\begin{array}{lll}
\delta u_b\delta T^{ab}_m\xi_a
-\delta h_b\delta\omega^{ab}\xi_a
&=&(u^c\xi_c)[h^2(\delta u)^2+(\delta h)^2]
+2[(\xi_c\delta u^c)(h_c\delta h^c)\\
&&-(h^c\xi_c)(\delta u_c\delta h^c)]
-2(\xi_c\delta h^c)\delta c\equiv Q_0.
\end{array}
\label{id}
\end{eqnarray}
If $\delta h^b$ varies freely, disregarding
$\delta c_\lambda=0$,
the quadratic $Q_0$ fails to be positive definite.

In the presence of the constraint variations with respect to
the $c_\lambda=0$,
the last term in the expression for $Q_0$ in (\ref{id})
vanishes identically:
\begin{eqnarray}
-2(\xi_c\delta h^c)\delta c=\sum_\lambda
-2(\xi_c\delta h^c_\lambda)\delta c_\lambda\equiv0,
\end{eqnarray}
and we are left with
\begin{eqnarray}
\begin{array}{ll}
Q_m:&=
\delta u_b\delta T^{ab}_m\xi_a
-\delta h_b\delta\omega^{ab}\xi_a\\
&=(u^c\xi_c)[h^2(\delta u)^2+(\delta h)^2]
+2[(\xi_c\delta u^c)(h_c\delta h^c)
-(h^c\xi_c)(\delta u_c\delta h^c)].
\end{array}
\end{eqnarray}
This expression $is$ (constraint) positive definite in
constraint variations with $\delta h\ne0$.
To see this, it is convenient to work in a local Lorenzian frame
of reference $g^{ab}=\eta^{ab}=\mbox{dia}(-1,1,1,1)$ with
$\{x^a\}=(x^0,x^\alpha)$, $\alpha=1,2,3$, such that
$u^b=(1,0,0,0)$. The magnetic field, $h^b_\lambda$, and
the variations $\delta u^b$ and $\delta h^b_\lambda$
may then be expressed in terms of three-component, spatial
vectors,
$k^\beta_\lambda$,
$\delta v^\beta$ and $\delta k^\beta$, respectively, as
\begin{eqnarray}
\begin{array}{l}
h^b_\lambda=k^\beta_\lambda\eta_\beta^b,\\
\delta u^b=\delta v^\beta\eta_\beta^b,\\
\delta h^b_\lambda=-u^b\delta\epsilon_\lambda
+\delta k^\beta_\lambda\eta_\beta^b.
\end{array}
\end{eqnarray}
Notice that $\delta \epsilon_\lambda$ follows from
$\delta c_\lambda=\delta\epsilon_\lambda+
k^\mu_\lambda\delta v_\mu=0$.
In the constraint variation of $\delta$,
the right hand-side in the quadratic variation in (\ref{id})
becomes a $\lambda-$sum of quadratic forms
$\delta W_{A\lambda}Q^{AB}_\lambda \delta W_{B\lambda}$
in $\delta W_{A \lambda}=(\delta v_\alpha,\delta k_{\alpha\lambda})$, where
\begin{eqnarray}
Q^{AB}_\lambda
=\left(
\begin{array}{cc}
 (u^c\xi_c)[k^2\eta^{\alpha\beta}-k^\alpha_\lambda k^\beta_\lambda]
& [\xi^\alpha k^\beta_\lambda-(k^\mu_\lambda\xi_\mu)\eta^{\alpha\beta}]\\
 {[}k^\alpha_\lambda\xi^\beta-(k^\mu_\lambda\xi_\mu)\eta^{\alpha\beta}]
&(u^c\xi_c)[\eta^{\alpha\beta}]
\end{array}
\right).
\end{eqnarray}
With $(u^c\xi_c)=\xi_0>0$,
the six eigenvalues, $s_l$, of $Q^{AB}_\lambda$ may be evaluated as
\begin{eqnarray}
\left\{
\begin{array}{l}
s_1=0,\\
s_2=(u^c\xi_c)>0,\\
s_3^\pm=
A\pm(A^2-[(u^c\xi_c)^2k^2-(k^c\xi_c)^2])^{\frac{1}{2}}>0,\\
s_4^\pm=
A\pm(A^2-k^2|\xi^2|)^{\frac{1}{2}}>0,
\end{array}
\right.
\end{eqnarray}
where $A=\frac{1}{2}(u^c\xi_c)(k^2+1)$.
Here, we have used $\xi^2=-\xi_0^2+\xi^\alpha\xi_\alpha<0$, so that
$\xi_0^2=|\xi^2|+\xi^\alpha\xi_\alpha\ge|\xi^2|$
and $\xi_0^2>\xi_\alpha^2$ for each $\alpha=1,2,3$.
Notice that the eigenvector
of $Q^{AB}_\lambda$ associated
with $s_1=0$ is $(k^\beta,0)$.
Therefore,
$W_{A\lambda}Q^{AB}_\lambda W_{B\lambda}$ is positive semi-definite,
and positive definite whenever $\delta k^b\ne0$.
The sum
\begin{eqnarray}
\sum_\lambda W_{A\lambda}Q^{AB}_\lambda W_{B\lambda},
\end{eqnarray}
therefore, naturally shares the same properties.

If $Q_f$
is positive definite with respect to $\delta V^f_A$,
the preceeding result shows that
\begin{eqnarray}
\delta V_A\delta U^A=Q_f+Q_m
\end{eqnarray}
is constraint positive definite with respect to
the $c_\lambda=0$. $\Box$

The constraint symmetrization procedure from Section 5.2
may now be applied to ideal Yang-Mills in divergence form
with $V_A$ as given above. It remains to ascertain
the orthogonality requirement
\begin{eqnarray}
\begin{array}{ll}
W_Af^A&=h_cf^c_m=-h^\lambda_cc_{\lambda\mu\nu}A^\mu_e\omega^{\nu e c}\\
&=(u^ch_c^\lambda)c_{\lambda\mu\nu}h^{\nu e}A^{\mu}_e
-(h^\lambda_ch^{\nu c})c_{\lambda\mu\nu}(A^\mu_eu^e)\\
&=c^\lambda(h^{\nu e})A_e^\mu c_{\lambda\mu\nu}=0,
\end{array}
\end{eqnarray}
where we have used total anitsymmetry of the structure
constants.
This completes our symmetrization
procedure for ideal Yang-Mills fluids.

In closing, we remark that the simple structure of the
flux-freezing constraints
allows for explicit
three-parametrizations
\begin{eqnarray}
h^b_\lambda=
-\frac{(k^c_\lambda u_c)}{(\xi^cu_c)}\xi^b+k_\lambda^b
\end{eqnarray}
for each Lie-component of the magnetic field, where
the three-parametrized Lie-component $k_\lambda$ satisfies
$k_\lambda^c\xi_c\equiv0$, which are such that
$c_\lambda\equiv0$.
By suitable choice of coordinates,
$\xi^b=(1,0,0,0)$
so that $k_a=(0,k_\alpha)$.
Our foregoing analysis shows that in the
$5+3N$ system parametrization
$V^\prime=(v_\alpha,k_\alpha,f,T)$ is such that
Friedrichs' Properties CI and CII are satisfied
(with the same $W_A=(v_\alpha,h_a,f-TS,T)$ as before,
but now in this
reduced parametrization and
with respect to all variations in this parametrization).
Thus, in the case of such simple constraints,
we could resort to pursuing Friedrichs' symmetrization
procedure. We are then left with the task of proving that
the symmetric hyperbolic system as defined by
Friedrichs symmetrization procedure actually solves
the entire system, now overdetermined by $N+1$.
Further dependency relations must be invoked, namely those
resulting from antisymmetry of $h_{[a}u_{b]}$.
We have not set out to pursue this, to see if
such specific treatment results in
shorter arguments,
although this should lead to a successful argument as well.
\mbox{}\\
{\bf Acknowledgement.} The author is very greatful to D.M. Eardley
and M. Stone for their support.
This work has been in response to Y. Choquet-Bruhat's
question on the characteristic determinant of IYMF, and the
author wishes to acknowledge
her critical comments on the early part of this work.


\bibliography{secbib}
\end{document}